\pdfoutput=1
\documentclass[aps,pre,twocolumn,superscriptaddress,showkeys,a4paper]{revtex4}
\usepackage{dcolumn}%
\usepackage{bm}%
\usepackage{graphicx}%
\usepackage{color}
\usepackage{subfigure}
\usepackage{hyperref}
\usepackage{latexsym}
\usepackage{amsthm}
\usepackage{amssymb}

\definecolor{MyColor}{rgb}{0.0,0.5,0.5}
\definecolor{MyColorNota}{rgb}{0.9,0.2,0.2}

\usepackage{soul}
\setulcolor{MyColorNota}
\usepackage{pifont}

\usepackage{framed}


\begin{document}
\title{Detection of timescales  in evolving complex systems}

\author{Richard K. Darst}
\affiliation{Department of Computer Science, Aalto University School
  of Science, P.O.  Box 15400,  FI-00076, Finland}
\author{Clara Granell}
\affiliation{Departament d'Enginyeria Inform\`atica i Matem\`atiques,
  Universitat Rovira i Virgili, 43007 Tarragona, Spain}
\author{Alex Arenas}
\affiliation{Departament d'Enginyeria Inform\`atica i Matem\`atiques,
  Universitat Rovira i Virgili, 43007 Tarragona, Spain}
\author{Sergio G\'omez}
\affiliation{Departament d'Enginyeria Inform\`atica i Matem\`atiques,
  Universitat Rovira i Virgili, 43007 Tarragona, Spain}
\author{Jari Saram\"aki}
\affiliation{Department of Computer Science, Aalto University School
  of Science, P.O.  Box 15400,  FI-00076, Finland}
\author{Santo Fortunato}
\affiliation{Department of Computer Science, Aalto University School
  of Science, P.O.  Box 15400,  FI-00076, Finland}
\affiliation{Center for Complex Networks and Systems, School of
  Informatics and Computing, Indiana University, Bloomington, IN, USA}

\begin{abstract}
  Most complex systems are intrinsically dynamic in nature. 
The evolution of a dynamic complex system is typically represented as a
  sequence of snapshots, where each snapshot describes the configuration of the 
  system at a particular instant of time.
Then, one may directly follow how the snapshots evolve in time, or 
aggregate the snapshots within some time intervals to form representative "slices" of the evolution of the system configuration. This is often
  done with constant intervals, whose duration is based on arguments on the nature of the system and of its dynamics. 
  A more refined approach would be to consider the rate of activity in the system
  to perform a separation of timescales. However, an even better alternative would be to define dynamic intervals
  that match the evolution of the system's configuration. To this end, we propose a method that aims 
  at detecting evolutionary changes in the configuration of a complex system, and generates intervals accordingly.
  We show that evolutionary timescales can be identified by looking for peaks in the similarity between the sets of events on
  consecutive time intervals of data.
  Tests on simple toy models reveal that
  the technique is able to detect evolutionary timescales of time-varying data both when the evolution is
  smooth as well as when it changes sharply. This is further
  corroborated by analyses of several real datasets.  Our method is
  scalable to extremely large datasets and is computationally
  efficient.  This allows a quick, parameter-free detection of multiple
  timescales in the evolution of a complex system.
\end{abstract}

\maketitle

We live in a dynamic world, where most things are subject to
steady change. Whether we consider the interactions between people,
proteins, or Internet devices, there is a complex dynamics that may
progress continuously with varying rates~\cite{saramakimoro2015}, or be punctuated by sudden
bursts~\cite{barabasi10}. Therefore, for understanding such complex
systems, rich empirical data with detailed time-domain information is
required, together with proper statistical tools to make use of
it. Fortunately, thanks to Information and Communication Technologies
(ICT) and the Web, time-stamped datasets are increasingly
available. However, the development of methods for extracting useful
information out of massive temporal data sets is still an ongoing task
(see e.g.\ \cite{holme12}).

A typical approach for studying the dynamics of a large-scale evolving complex
system is to divide the temporal evolution into meaningful intervals where information of single snapshots is aggregated.
The ``slice'' corresponding to the
interval between times $t_n$ and $t_{n+1}$ then comprises the subset
of elements and interactions (or events) that are active between $t_n$ and
$t_{n+1}$. For instance, a slice may correspond to a group of
people exchanging phone calls during an interval of one hour, or to a set of
tweets containing specific hashtags posted within some time span. Once
the intervals are defined, the analysis of the system turns into
an investigation of the slices.

The main problem is then how to properly ``slice''  the data, so that the resulting slices provide the understanding we are after.
Among several alternatives, one can decide to use a constant slice size corresponding to some characteristic temporal scale of the dynamics (if there is any), or
a variable slice size that follows the rate of the dynamic activity. However, if we focus our attention on the evolutionary aspects of the system,
it might be convenient to define the slices according to the variability of the system through time.  As an example, if we want to use data on email exchanges within the company to understand the evolution of the Enron scandal, it is better to use slices that capture changes in the composition of communication groups and thus track the evolution of communication patterns, as compared to slices based solely on the rate of activity in the email communication network.
Consequently, the choice of slice size will be determined
both by the availability and granularity of data, and by the
timescales that reflect evolutionary changes in the configuration of events.

It follows that there is a clear need for a principled way of
automatically identifying suitable evolutionary timescales in data-driven
investigations of evolving complex systems. This is the motivation behind the
method proposed in the present paper. A proper
time-slicing method should produce meaningful evolutionary timescales (intervals)
describing changes in the event landscape, both when these changes are smooth and when they are abrupt.
For abrupt changes, the problem becomes related to
{\it anomaly detection}~\cite{chandola09,akoglu14} that has
applications in e.g.\ fraud or malware detection, and in particular to
{\it change-point detection} in temporal networks that comprise
time-stamped contact events between nodes.
For temporal networks,
several ways for detecting change points have been proposed, based on
techniques of statistical quality
control~\cite{lindquist07,mcculloh11}, generative network
models~\cite{peel14,peixoto15b,peixoto15c}, bootstrapping~\cite{barnett14}, and
snapshot clustering~\cite{tong08b,berlingerio10,berlingerio13}.
Note that our approach is not specifically designed for change-point detection, but 
it does reveal change points as a side product of the timescale detection.

In this work, we introduce an automatic time-slicing method for detecting timescales in the evolution of complex systems that consist of recurring events, such as recurring phone calls between friends, emails between
colleagues and coworkers, mobility patterns of a set of individuals, or Twitter
activity about a certain topic.
The method can be easily extended to systems in which interactions are permanent after
their appearance (e.g.\ citation networks).  Our approach is sequential, and consists of determining the size of each interval from
the size of the previous interval, trying to maximize the similarity between
the sets of events within consecutive intervals (slices). In general, there is a unique maximum:
too short intervals contain few, random events whereas too long intervals are no longer representative of the system's state at any point. Consider, e.g., a set of phone calls in a social system aggregated for a month -- compared to this slice, one-minute slices would contain apparently random calls, and a 10-year slice would contain system configurations that have nothing to do with the initial slice.
The method is parameter-free and validated on toy models and real datasets.

\section{Results}

\subsection{Method for detecting evolutionary timescales}

\label{sec-segm}

Let us consider a series of time-stamped events recorded during some period of time. %
Our goal is to create time slices, i.e.\ sets of consecutive time intervals, that capture the significant evolutionary timescales
of the dynamics of the system.  There are some minimal conditions that data must fulfil for this to be a useful procedure: events must be recurrent in time and the total period of recording must be long enough to capture the evolution of the system.

In this framework, the appropriate timescales detected must have the following properties:
i) Evolution is understood as changes in the event landscape, i.e., intervals will only depend on the identities of events within them, not on
the rate of events in time. ii) Timescales should become shorter when the
system is evolving faster, and longer when it is evolving more slowly.
iii) The particular times when all events suddenly change (critical times) should have
an interval placed at exactly that point. This is a limiting case of (ii).

To obtain timescales satisfying the above properties, our method iteratively constructs the time
intervals $[t_n,t_{n+1})$ according to a similarity measure between
each pair of consecutive slices $S_{n-1}$ and $S_n$, where $S_n$ is
the set of events between time $t_n$ and $t_{n+1}$.
By looking at the similarity measure as a function of time, one can
detect multiple time scales.  The algorithm is designed to detect the
longest possible timescales in a parameter-free fashion.  For
discussion of timescale detection and why this is a good choice for
real data, see Sec.~A.1.

For the sake of clarity, assume that our method is in progress and we have a previous slice
$S_{n-1}$.  We want to find the next slice $S_{n}$ that is defined
by the interval $[t_{n}, t_{n}+\Delta t_{n})$. The increment
$\Delta t_{n}$ is the objective variable, and is determined by
locally optimizing the Jaccard index between $S_{n-1}$ and $S_{n}$:
\begin{equation}
  J(\Delta t_{n}) = \frac{S_{n-1} \cap S_{n}(\Delta t_{n})}{S_{n-1} \cup S_{n}(\Delta t_{n})}.
\label{jac}
\end{equation}
This is shown in Fig.~\ref{fig:schematic}(a).  The Jaccard
Index \cite{jaccard01} $J(\Delta t_{n})$ estimates the similarity of
the two sets as the fraction of their common items (events) with respect to
the total number of different items in them.  In
our case, the optimization process is performed by increasing $\Delta
t_{n}$ until a peak for $J(\Delta t_{n})$ is found (see
Materials and Methods). The Jaccard similarity can then be used to
understand the underlying event dynamics.

This iterative process requires an initial slice, which is
computed as follows: Beginning with the initial time $t_0$, we construct two
intervals $[t_0, t_0+\Delta t)$ and $[t_0+\Delta t, t_0+2\Delta t)$.
Repeating the process above, we find the maximum of $J(\Delta t)$ and
use that to set the first two intervals at once. This initialization
preserves the meaning of our slices and does not include any
additional hypothesis.  After computing both slices, the second one is
discarded so that all subsequent intervals have the same semantic meaning, and
we start the iterative process described above.
Note that the method is parameter free. The Jaccard Index is used in a similar way in \cite{berlingerio10,berlingerio13} to
assess the similarity of consecutive snapshots that are eventually grouped
by means of hierarchical clustering to build relevant intervals, and in~\cite{krings12} to study the effects of time window lengths in social network analysis.

The logic behind the method is easy to understand.  Let us first
consider a smoothly evolving system. We take the
previous interval $S_{n-1}$ as fixed, and try to find the
next interval $S_n(\Delta t)$.  On one hand, if $\Delta t$ is
small, increasing $\Delta t$ will very likely add events already present in
$S_{n-1}$.  Thus, the numerator (intersection term) of Eq.(\ref{jac})
increases more than the denominator (union term), and $J(\Delta
t)$ becomes larger.  On the other hand, if $\Delta t$ is very large, the increase in
$\Delta t$ will tend to add new events not seen in the previous interval. Therefore the
denominator (union term) of Eq.(\ref{jac}) grows faster than the
numerator (intersection term), and consequently $J(\Delta t)$
decreases.  These principles combined imply the existence of an
intermediate value of $\Delta t$ representing a balance of
these factors, which is a solution satisfying our
basic requirements of sliced data.  If the smoothness in the evolution
of the temporal data is altered by an abrupt change (critical time), the previous
reasoning still holds, and a slice boundary will be placed directly at the snapshot where the anomalous behavior begins.

Our method has several technical advantages.  First and foremost, it is fast and scalable.
The method is $O(N)$ in the total data size $N$ (e.g.\ number of distinct events), if intervals are small.
Efficient hash table set implementations also allow linear scaling in
interval size, and therefore extremely large data sizes can be processed.
The data processing is done online: only the events from the previous interval must be saved to compute the next interval.
There are no input parameters or \textit{a priori} assumptions that must be made before the method can begin.  The input is simple, consisting simply of (event\_id, time) tuples.
The method finds an initial intrinsic scale to the data, where each interval
represents roughly the same amount of change.

\subsection{Validation on synthetic data}

\begin{figure}
  \centering
  \includegraphics[width=\columnwidth]{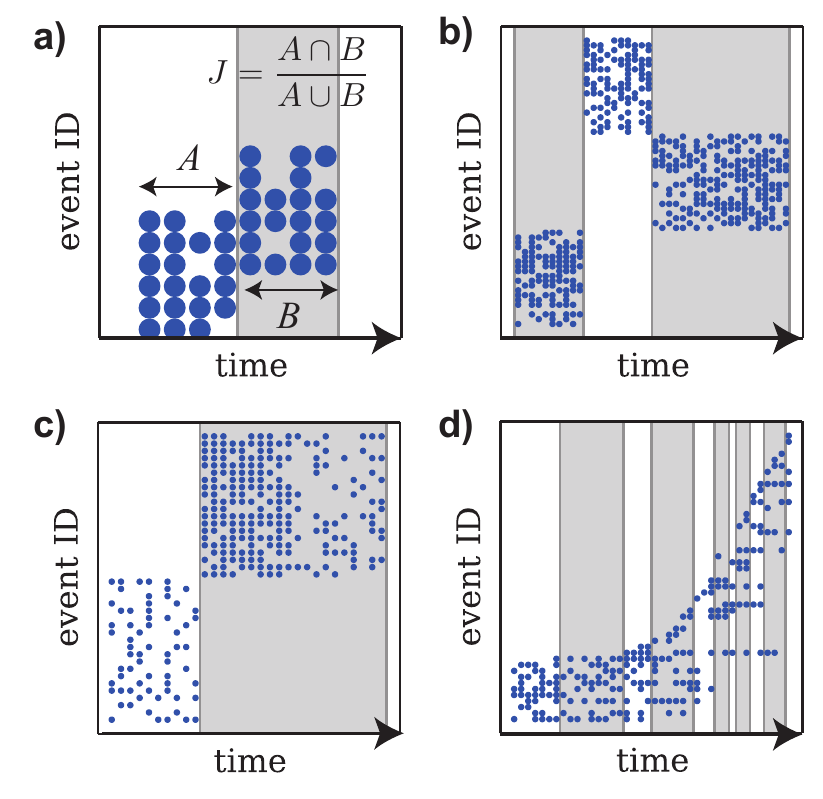}
  \caption{Schematic representation of the application of our method to synthetic data.  In
    each figure, the horizontal axis represents time in arbitrary units,
    and the vertical axis the event ID.  A dot is present at each
    point where a given event occurs at a given time. The gray and white
    segments show the detected intervals.
    (a) The Jaccard score $J$ is computed
    for the sets of events between adjacent intervals.  Intervals are
    adjusted to maximize $J$.
    (b) Interval identification in a trivial case.
    Each region contains a characteristic set of events, all distinct from those of the other regions.
    (c) The total event rate (see second interval) does
    not affect interval size, only the characteristic set of events does.
    (d) Example of
    varying interval sizes.  In the first half of time, events
    change more slowly than in the second half, and thus intervals are
    longer.  Because this is data with clear transitions, (a-c) have
    first intervals merged.}
  \label{fig:schematic}
\end{figure}

First we have generated synthetic temporal data examples to be analyzed with our method.
The data displays several common configurations of the temporal evolution of a toy system. Fig.~\ref{fig:schematic}
shows some examples and the timescales that our method detects.
In Fig.~\ref{fig:schematic}~(a) we sketch the details of the method.
Fig.~\ref{fig:schematic}~(b) shows three clearly distinguishable sets of events,
which are correctly sliced into their respective timescales.
Within each interval, events repeat
very often, but between intervals there are no repetitions.
Fig.~\ref{fig:schematic}~(c) shows that the rate of the short-term
repetitions does not matter from our evolutionary point of view.
The second interval has the same
characteristic events at all times, so no evolutionary changes are observed.
Fig.~\ref{fig:schematic}~(d) shows a continually evolving system. For the first half, 
the long-term rate of evolutionary change is slow (events are 
repeated for a longer time before dying out), and thus the intervals are
longer.  Thereafter, the process occurs much faster, and thus
the intervals are shorter.

We have also validated our method on more complex synthetic data
designed to reveal non-trivial but controlled evolutionary timescales.
We choose a dynamics where there is a certain fixed number of events
which, at every time, might be active or inactive (short-term
repetition). We impose that the volume of active events per time unit
must be, on average, constant. Additionally, to account for the
long-term evolution, we change the identities of the events at a
(periodically) varying rate. Low rates of change, that is, changing the 
identities of only a few events at any time, should result in large time intervals for the
slices, while faster rates lead to high variations in the identities
of the events and should produce shorter slices. %

\begin{figure*}
  \centering
      \includegraphics[width=2\columnwidth]{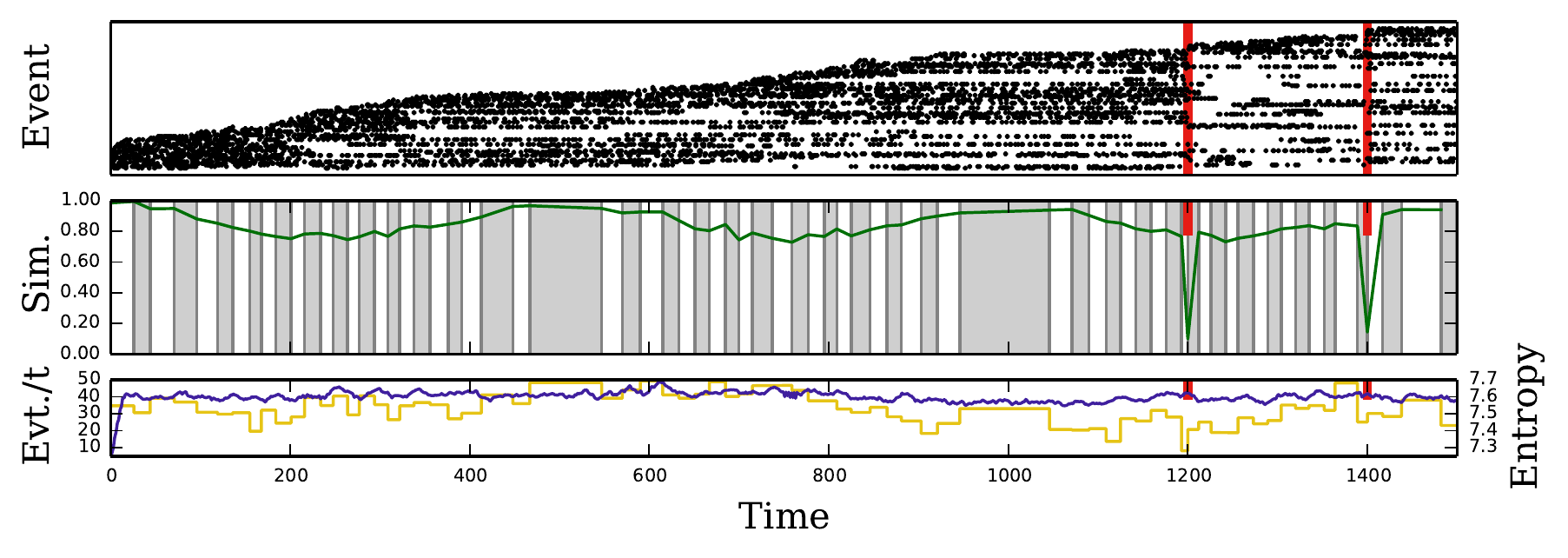}
      \caption{Artificial data with changing event turnover rate with period
        $\tau=500$.
        (Top) Representation of events.  Events have been ordered
        to show the envelope of the dynamics.  Events change
        at the slowest rate at $t=500,1000$, and there are critical
        points at $t=1200,1400$.  For clarity, only 10\% of events are shown.
        (Middle) Detected intervals (gray/white regions),
        and inter-interval similarity (plotted green lines)
        for the slicing algorithm.  We can see that the internal length
        reacts to the rate of change of events, and that the critical
        points correspond to intervals boundaries and to
        large drops in similarity.  The similarity also reflects the
        rate of change of events at other times.
        (Bottom) Event rate (dark blue) and interval entropy (light
        yellow) over time.  We see that overall event rate is
        constant and cannot be used to detect dynamical properties.
        We also see that each interval contains about the
        same amount of information, 7.3 to 7.7 bits.}
  \label{fig:critical}
\end{figure*}

We check our method on this benchmark using a periodic evolutionary rate, with
period $\tau=500$ (see Materials and Methods). In Fig.~\ref{fig:critical}, the method shows the
desired behavior, namely, it produces short intervals in regions where change
is fastest ($t\approx 250, 750$) and longer intervals for low variability
of events, i.e.\ slow evolution ($t\approx 0, 500, 1000$).  Most notably, the
Jaccard similarity is seen to reflect these dynamics, too.  When
change is slowest, the inter-interval similarity is high, and vice
versa.  These similarities could be used to perform some form of
agglomerative clustering to produce super-intervals representing longer-term dynamics.  
We also show the Shannon entropy
within each interval.  We see that our method produces intervals
containing roughly the same amount of information.

At times $t=1200$ and $t=1400$, we add two ``critical points'', when
the entire set of active events changes instantly.  This is the limiting case
of evolutionary change.  As one
would expect, slice boundaries are placed at exactly these points.
Furthermore, the drastic drop in similarity indicates that unique change
has happened here, and that dynamics across this boundary are uncorrelated.

\subsection{Analysis of real temporal data}

In this section we apply our approach to real-world data, where any true signal is obfuscated by
noise and we do not have any information about the underlying dynamics.

We start with the Enron email dataset (see Materials and Methods).
This dataset was made public during the legal investigation concerning the Enron Corporation, which ended in the bankruptcy and collapse of this corporation~\cite{klimt2004enron}.
Here, each event is an email communication sent or received by any of the senior managers who were subsequently investigated.
During the course of the investigation, there were major structural changes in the company (see Appendix B).
Fig.~\ref{fig:enron-pvalue} shows the results of applying the method to the Enron data. Generally, the intervals are of the order of several weeks, which appears to be a reasonable time frame for changes in email communication patterns. 
As expected, we see that the detected intervals do not simply follow changes in the rate of events, but rather in event composition.
When comparing interval boundaries to some of major events of the scandal (exogenous information, shown as red vertical lines), we see that the interval boundaries often align with the events; note that there is no a priori reason for external events always to result in abrupt changes in communication patterns. 
Additionally, we present the results of the application of the change point
detection method of Peel and Clauset (\textit{cpdetect})\cite{peel14}. Unlike \textit{cpdetect},
our method reveals the underlying basic evolutionary timescale (weeks), rather than
only major change points.  On the other hand, \textit{cpdetect} requires a
basic scale as an input before its application (in this case, one week).  Performing a
randomization hypothesis test, we find evidence (at significance $p=0.0194$)
that \textit{cpdetect} change points correspond to lower-than-random similarity values
within our method (See Appendix B).

\begin{figure*}
  \centering
  \includegraphics[width=2\columnwidth]{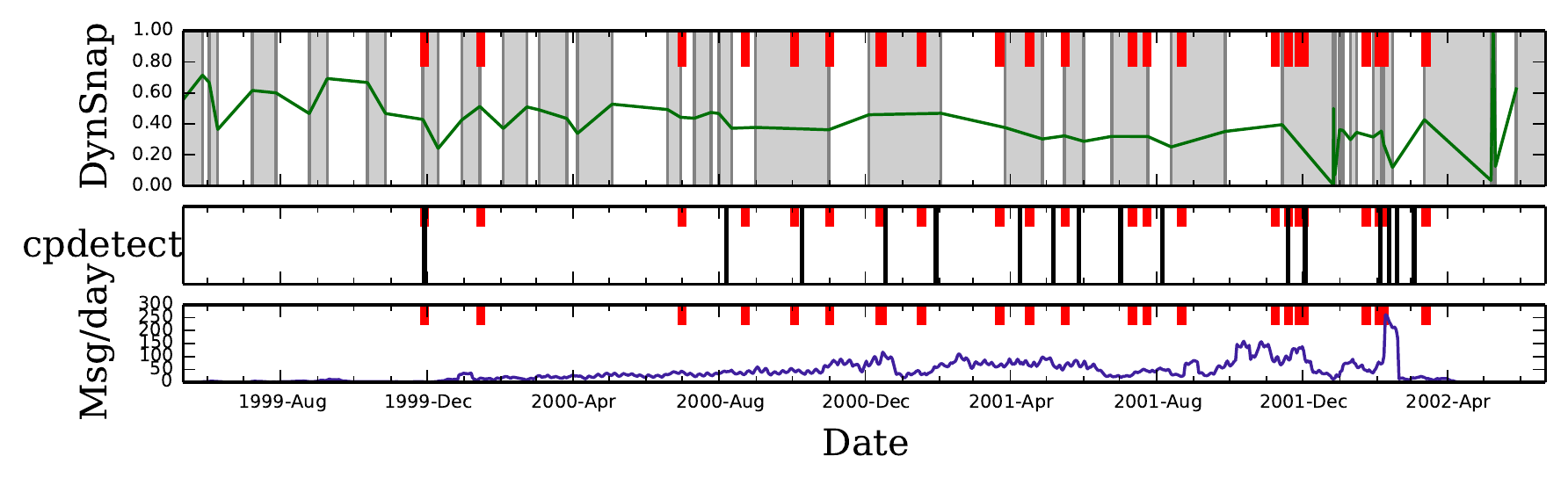} \\
  \caption{Comparison of our method and the change point detection
    method of Ref.~\cite{peel14} on the Enron email data.  Thick red lines
    indicate events from the company's timeline.
    (Top) Results of our method on the
    network, in the same format as Fig.~\ref{fig:critical}.
    (Middle)
    Results of the \textit{cpdetect} method on the network.  Dark vertical lines
    mark change points.
    (Bottom) Event density of the network.
    We see that both methods align with the company's events to
    some degree.  Our method provides more fine-grained intervals,
    not just the major points, and is capable of more accurately
    aligning with the real events since it does not have predefined
    window sizes as an input.}
  \label{fig:enron-pvalue}
\end{figure*}

\begin{figure*}
  \includegraphics[width=2\columnwidth]{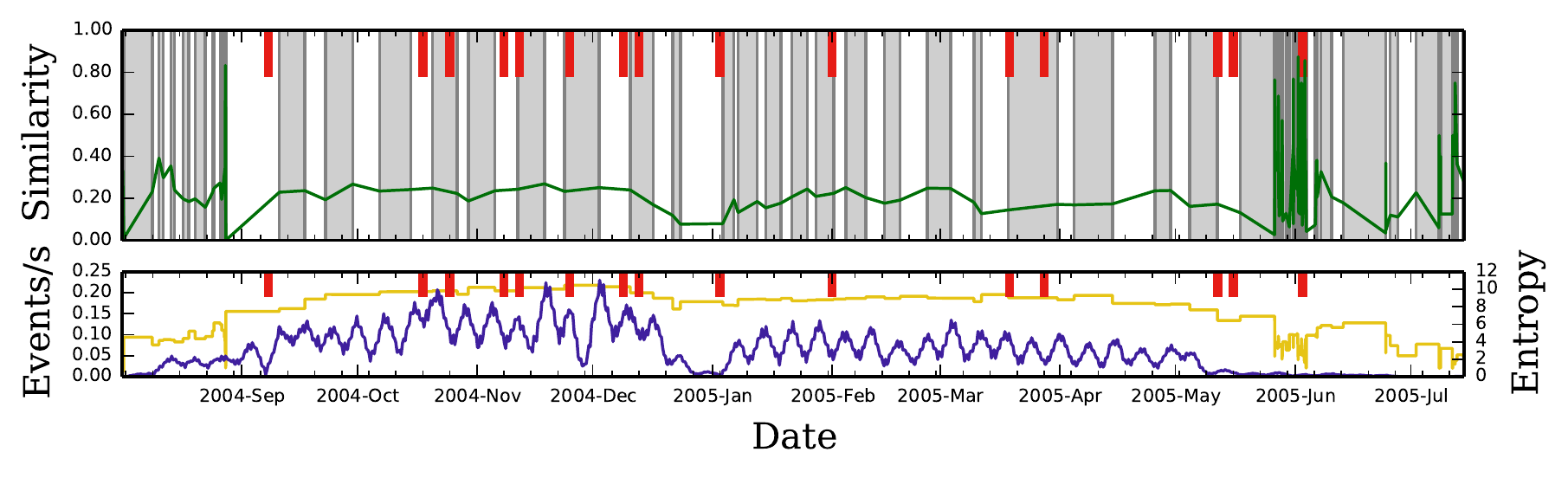}
  \caption{Slicing of the MIT reality mining dataset.
    In this dataset, an event is considered the (from, to) pair of
    Bluetooth scanning, representing two devices being located within
    proximity of each other.
    The times of major events are known, such as important
    conferences or start/ends of semesters (see Appendix B3). We see that the
    slicing can find such major events, including the
    beginning of the 2005 semester and the beginning of classes at
    2005-Feb.}
  \label{fig:reality}
\end{figure*}

Next, we focus on the MIT Reality Mining personal mobility dataset.
In this experiment, about 100 subjects, most of them affiliated to MIT Media Lab, have cellphones which periodically scan for nearby devices using the Bluetooth
personal area network~\cite{eagle2006reality}. Here, events correspond to pairs of devices being in Bluetooth range (i.e.\ physically close).
Note that we do not limit the analysis to interactions only between the 100 subjects, but process all $\approx 5 \times 10^4$  unique Bluetooth pairs seen (including detected devices not part of the experiment), a total of $1.8 \times 10^6$ distinct data points.
The major events (see Appendix B3) correspond to particular days in
the MIT academic calendar or internal Media Lab events.
In Fig.~\ref{fig:reality} we plot the time slices generated by our method.
Again, we see that the basic evolutionary timescale is about one week.
We see a correlation with some of the real events, such as a long period during winter holidays (late December) and Spring Break (late March).
At the beginning (August) and end (June) there is a minimal data volume, assimilable to noise, which precludes the method to find reliable timescales.

In our last example, we look at the social media response to the
semifinals of the UEFA Champions League 2014-2015, a major European soccer tournament.
We collected data from Twitter looking for tweets containing the hashtag \texttt{\#UCL}, the most widely used hashtag to refer to the competition.
As all captured tweets contain the hashtag \texttt{\#UCL}, we define events by considering the other hashtags that a tweet may contain.

\begin{figure*}
  \includegraphics[width=2\columnwidth]{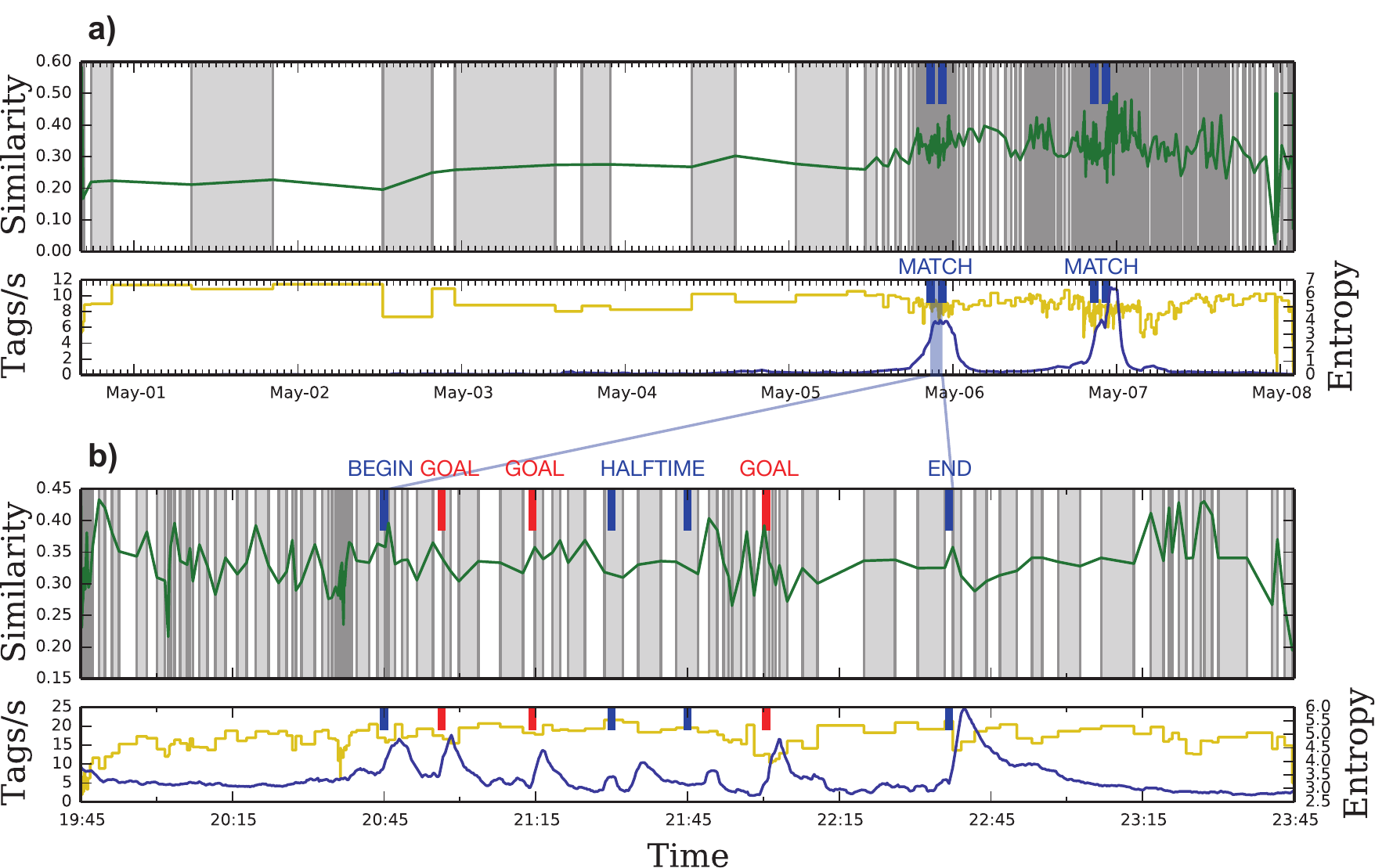}
  \caption{a) Slicing of Twitter hashtag sequences generated during UEFA Champions
    League 2014-2015 football tournament. The format of the figure is as in
    Fig.~\ref{fig:critical}.
    In this data all tweets corresponding to the
    \texttt{\#UCL} hashtag are captured, so the rest of the co-occurring hashtags
    are taken as our events. The two increments in the event rate represent the two matches,
    which are captured through very fine intervals in the data slicing. b) First semifinal of UEFA Champions League 2014-2015, between Juventus and Real Madrid.  The format
    of the figure is as in Fig.~\ref{fig:critical}.  The
    game runs from 20:45 until 22:30, with the halftime from 21:30
    until 21:45.  We see sped-up activity around the goals in the
    slicing (top) that is not simply explained by the
    event rate (bottom).  When goals are scored %
    (20:53, 21:12 and 21:55)
    there is a much faster topics turnover.}
  \label{fig:hashtags1}
\end{figure*}

If a tweet contains more than one other hashtag, each of them is considered as a separate event. For example, if at time $t$ there is a tweet containing \texttt{\#UCL, \#Goal} and \texttt{\#Football}, we consider that we have two events at time $t$, one for \texttt{\#Goal} and the other for \texttt{\#Football}.
In Fig.~\ref{fig:hashtags1}, we see a one-week span around the two first semifinal games.
Because Twitter discourse is rather unstructured, intervals are very small, comprising at most a few hours.
During the games themselves (May 5, 6 evenings), we see extremely short intervals representing the fast evolution of topics.
In Fig.~\ref{fig:hashtags1}, we zoom in the first semifinal game.
As the game begins, intervals become much shorter as we capture events that did not manifest in the previous games to the match.
We see other useful signatures, like the rapid dynamics at the beginning of the second half (21:45) and at the time of the game-winning goal (21:58).

\section{Discussion}

Complex systems are inherently dynamic on varying timescales. Given the amount of available data on such systems, especially on human behavior and dynamics, there is a need for fast and scalable methods for gaining insight on their evolution. Here, we have presented a method for automatically slicing the time evolution of a complex system to intervals (evolutionary timescales) describing changes in the event landscape, so that the evolution of the system may be studied with help of the resulting sequence of slices.

Our method provides the starting point of
a pipeline analysis.  Many time-varying data analysis techniques
require segmented data as an input, and if the intervals are decided according to some fixed
divisions, the boundaries will not be in the optimal places.  Our
method places snapshot boundaries in principled locations, which will
tend to be the places with the highest rate of change - or actual
sharp boundaries, if they exist.  After this process, the sliced data
can be input to other methods which further aggregate them into even
higher orders of structure while respecting the underlying dynamics.

Our code for this method is available at \cite{rkd_dynsnap_code} and
released under the GNU General Public License.

\section{Materials and Methods}
\textit{Identification of optimum interval}
The value of $\Delta t$ is searched from the raw data, and various strategies
can be used.  A simple linear search suffices, but is inefficient at
long timescales.  An event-driven approach checks only $\Delta t$
intervals corresponding to actual events, but this too is inefficient
at long timescales.
An ideal approach, and the one adopted in this
paper, is to use exponentially spaced intervals (See Appendix).  The shortest $\Delta
t$ searched is taken from the next event in the data.

\subsection{Temporal benchmark}
The proposed benchmark has a periodic activity change in time, while having a
uniform event rate.
A universe of $N=1000$ event IDs is created.  Upon initialization,
each event is independently placed into an activated status with
probability $q$.  At each time step, each active ID creates an event
with probability $p$.  Also, at each time step (and before event
creation), a fraction $c(t)$ of event IDs are picked.  Each of these
IDs has its active status updated, being made active with probability
$q$ regardless of its previous state.  This preserves the mean total
number of activated IDs as $qN$ at all times.  At each time step, on
average $pqN$ events are expressed.  Thus, by looking at simply the
rate of events, the model is completely uniform and there is a
constant average event rate in time.

The changeover rate $c(t)$ is periodic, however, obeying
\begin{equation}
  \label{eq:periodicct}
  c(t) = c_0 + c \left[\frac{1}{2} - \frac{1}{2} \cos\left({2\pi \frac{t}{\tau}}\right)\right],
\end{equation}
with $c_0$ being the minimum changeover rate, $c$ the scale
of changeover, and $\tau$ the period.
Thus, at some points, the IDs of the expressed
events is changing more rapidly than at other times, and this is the
cause of longer and shorter intervals.

When there is a critical event at time $t_\mathrm{crit}$, we
exceptionally set $c(t_\mathrm{crit})=1$ at that time.  This produces
a state of the system uncorrelated from the previous time.  According
to the model, this changeover occurs before the time step, which
matches with placing a new interval at $t_\mathrm{crit}$ in our
half-open segment convention.  However, the total universe of event
IDs stays the same, and there is no statically observable change in
behavior.

In this work, we use $N=1000$, $\tau=500$, $p=0.2$, $q=0.2$, $c_0=0$, and
$c=0.01$.

\subsection{Shannon Entropy}
The Shannon entropy \cite{shannon49} is calculated by
$H = -\sum p_i \ln p_i$
over a series of events $i$ with corresponding probability $p_i$.

\subsection{Enron temporal data}
The Enron Email Dataset consists of emails of approximately 150 senior
managers of the Enron Corporation, which collapsed in 2001 after
market manipulation was uncovered, leading to an accounting scandal
\cite{klimt2004enron}.  We have all bidirectional email communications
to and from each key person with a resolution of one day.  An event is
the unordered pair (source, destination) of each email.  Our
particular input data is already aggregated by day, therefore each
event is only repeated once each day, with a weight of the number of
mails that day.
Appendix Tables~I and~II lists the basic properties and
Appendix Table~III lists the major events of this dataset.

\subsection{Reality mining}
The Reality Mining dataset covers a group of persons affiliated with
the MIT Media Lab who were given phones which tracked other devices in
close proximity via the Bluetooth personal area network protocol.
\cite{eagle2006reality}.  An event is defined as every ordered pair
(personal\_device, other\_device).  We only have data from the
personal devices of the 91 subjects who completed the experiment, not
full ego networks.  The data contains 1881152 unique readings from
2004 August 8 until 2005 July 14, with most centered in the middle of
the academic year.
Appendix Table~IV lists basic
properties and Appendix Table~V lists the major events of this
dataset \cite{MITcalendar}.

\subsection{UCL hashtags}
In this dataset, we scrape Twitter for all hashtags containing
\texttt{\#UCL}, referring to the Europe Champions League tournament.
We scrape between the dates of 30 April 2015 and 08 May 2015, covering
the two first semifinal games of the tournament.  An event is any hashtag
co-occurring with \texttt{\#UCL}, except \texttt{\#UCL} itself.  If
there are multiple co-occurring hashtags, each counts as one event.
All hashtags are interpreted as UTF-8 Unicode and case-normalized
(converted to lowercase) before turning into events.
Appendix Table~VI lists basic properties and
Appendix Table~VII lists the major events.

\begin{acknowledgements}
  We thank Andrea Avena K\"onigsberger for useful discussions.
  R.K.D., A.A., S.G.\ and S.F.\ gratefully acknowledge MULTIPLEX, grant number
  317532 of the European Commission and the computational resources
  provided by Aalto University Science-IT project. R.K.D.\ and J.S.\ acknowledge Academy of Finland,
  grant number 260427. C.G., A.A.\ and S.G.\ acknowledge Spanish Ministerio de Econom\'{\i}a y
  Competitividad, grant number FIS2015-71582-C2-1. A.A.\ also acknowledges ICREA Academia and the
  James S.\ McDonnell Foundation.
\end{acknowledgements}

\bibliography{science,dynsnap,science2}

\appendix

\section{Algorithm description}

The algorithm for detecting the evolving timescales of time-varying data is implemented in different layers. %
Each layer is reasonably independent of others, allowing them to be improved or
replaced independently of each other. This allows the method
to be easily adapted and improved for new types of data. 
This document describes each of the possible variations of each layer of implementation.
Please note that this document is designed to explain the algorithm
components from a scientific viewpoint. It is not a usage manual of
the code, which is found on the project site. 

 The algorithm is designed in the following layers:

\begin{enumerate}
\item Data input (See user manual) %
\item Data representation (Sec.~\ref{sec:meth-representation})
\item Similarity measures (Sec.~\ref{sec:meth-similarity})
\item The core segmentation process (Sec.~\ref{sec:meth-time})
\item Choosing $\Delta t$ to test at one iteration (Sec.~\ref{sec:meth-dt})
\item Maximum finding (Sec.~\ref{sec:meth-peak})
\end{enumerate}

Code for the method is available at \url{https://github.com/rkdarst/dynsnap}.

\subsection{Data and timescales}
\label{sec:meth-prelim}
Before the method is applied to data, it is important to understand
the necessarily qualities of input data and how timescales are
detected.  The method is by default suited to most real data in a
parameter-free fashion, but there can be non-intuitive behavior for
data with long-term self similarity, such as completely periodic
artificial data.

The purpose of this method is to segment events into time-intervals so
that the spacing of intervals corresponds to the similarity of events
within those intervals.  This is captured by looking at the similarity
of events within intervals.  An interval is too short if there is not
enough time to get a characteristic set of events within it.  An
interval is too long if, by some definition, the start and end of the
interval are different in terms of characteristic events.  All else
being equal, we would rather take longer intervals.  Data must have two
properties in order to be sliced.  First, events must repeat on a
short time scale.  That is, events repeat in time, so that the same
event can be captured in two adjacent intervals, or else intervals can
not be similar.  This is our smallest possible slicing time.  If
events do not recur, then our method can not be used.  On the other
hand, there must be a long time scale at which events do \textit{not}
recur.  If events did always recur at every time scale, then the
longest time scale of self-similarity covers the entire time.
In this case, any equal-size slicing of the interval would be acceptable.
There
are two ways this long-term change can be understood.  One is if some
events become active, turn on and off for some time, and then
eventually deactivate.  That is, you can identify some time range when
an event occurs, and then forever before and after that time, that
event is not seen.  Alternatively, there can be a finite universe of
events, and the events are extremely bursty, so that there are
very long inactive periods between phases of activity.

Our method can detect different timescales.  If there are different
timescales, then $J(\Delta t)$ (Sec.~\ref{sec:meth-peak}) for each
slice will reveal them.  There is then the question of selecting which
of these timescales the algorithm should return for any given run of
the code.  The algorithm described here and the code released along
with this article adopts the strategy that, by default, the longest
possible timescale should be detected.  There are options to find
shorter timescales, and a detailed discussion exists within the code
manual at \texttt{doc/Manual.rst}, and in Sec.~\ref{sec:meth-peak} and
Sec.~\ref{sec:peak-factor}.

To use this method on data which does not meet the criteria above, in
particular the ``long term change'', timescale options will be
needed.  In particular, consider the case of periodically repeating
data, especially artificial data that has no random noise.  There is a
short term similarity timescale, smaller than the period of
repetition.  This is the most useful timescale to detect.  Detecting
this timescale would allow one to detect the components within the
repeating pattern.  However, there is also a similarity at a long term
timescale.  Detecting this timescale would override and hide all internal
similarity.  Since this similarity extends to all time and our method
by default tries to find the longest possible timescale, we would find
the (less interesting) result that all time is similar.  Another way
to see this is that there is as much, if not more, similarity between
the first 1/3rd and last 1/3rd of data as there is within each period.
(Note that this can be used as a test for data where time scale
adjustment is needed).  Thus, our method correctly detects the longest
timescale as covering all time.  However, note that any small amount
of non-repeating random noise will destroy this long-term similarity,
and the method then detects some shorter time scale.  This is why we
say that our method works well and without parameters on real data -
random noise destroys long-term similarity, so detecting the longest
time scale returns the expected result.

\subsection{Preliminaries}
\label{sec:meth-prelim}
All time ranges are considered half-open intervals $[t_1, t_2)$.
Within an interval, events are aggregated to produce ``sets of events'' %
which characterize the interval.  The core goal of the slicing algorithm is
to produce adjacent intervals that have different sets of events, but that are 
not too dissimilar.  %

\subsubsection{Data and interval representation}
\label{sec:meth-representation}
At the lowest level, all data is a multiset of \texttt{(time, id,
  weight)} tuples for each event. The variable \texttt{time} refers to the time at which an event has happened,
  \texttt{id} stands for the unique ID of the event, and \texttt{weight} stands for the number of occurrences of that 
  particular event at that particular time. This conforms a multiset because
duplicate events at the same time and ID are allowed. If data is unweighted, all weights
can be considered to be (and stored as) $1$.

When using data in an unweighted fashion, then the set of events is
a regular (non-weighted, non-multiset) set containing the IDs
of every event present within the interval.

When using data in a weighted fashion, a set of events is
a weighted set containing event IDs for every event present within the
interval, where each event has an associated (non-negative)
weight.  If the original data has only unit weights (or was originally
unweighted), then the event weights reduce to the counts of events
within the intervals. The weights are the sum of weights of all
events of that ID within that interval.

Note that a weighted set can be converted to unweighted by dropping all weights.
Conversely, if unweighted data is made weighted, the weights count the number of
events present. %

\subsection{Similarity measures}
\label{sec:meth-similarity}

The various similarity measures are defined as a function between two 
(possibly weighted) sets of events as defined above, with a resulting value in the
range $[0,1]$.  A $1$ similarity defines a perfect match while $0$
indicates no similarity.

In the following examples, consider two intervals $A$ and $B$.  Here we are
loose with terminology and use the terms $A$ and $B$ interchangeably to refer to the interval itself
as well as to the set of events within the intervals. %
 In the remainder of this document, we use $J$ to refer to a generic similarity measure,
even if the symbol itself refers to the Jaccard score.

\subsubsection{Unweighted Jaccard}
This measure calculates the similarity between unweighted sets of events. %
It makes use of the standard Jaccard score,
\begin{equation}
  \label{eq:jaccard}
  J(A,B) = \frac{|A \cap B|}{|A \cup B|}.
\end{equation}
In this formulation, the Jaccard score is $1$ if two intervals have the same elements
regardless of the number of occurrences of those elements within the intervals.

\subsubsection{Weighted Jaccard}
The following is the extension of the Jaccard score to the intersection
and union of weighted sets.  Weighted sets are defined by real-valued
indicator functions $w_{i}$ representing the weight of each element within
the set.  Element $i$ has a weight of $w_i$.  Any element of weight
zero is considered to not be contained in
the set and can be removed. Conversely, any element not present in the set has a weight of zero.
A weighted union is defined to have elements of
\begin{equation}
  \label{eq:union}
  w_{U,i} = \max(w_{A,i}, w_{B,i})
\end{equation}
over all elements in either $A$ or $B$.  Here, $w_{U,i}$ is the
indicator function for the union, and respectively $w_{A,i}$ and
$w_{B,i}$ for the sets $A$ and $B$.  A weighted intersection is
defined to have elements of
\begin{equation}
  w_{I,i} = \min(w_{A,i}, w_{B,i})
\end{equation}
with components analogous to Eq.~(\ref{eq:union}).  With these
definitions for the intersection and union, the weighted Jaccard score
is computed as in Eq.~(\ref{eq:jaccard}).

It is worth noting that the weighted Jaccard score introduces a bias towards equal-size sets
with equal element counts.  Thus, there is some ``inertia'' in
interval sizes and can not adapt to changing timescale quickly.

\subsubsection{Cosine similarity}
The weighted sets can be considered sparse vectors,
allowing us to use the cosine similarity.  Defined in terms of sets, the
cosine similarity is
\begin{equation}
  C(A,B)
    = \frac{ A \cdot B }{ |A| |B| }
    = \frac{ \sum(w_{A,i}  w_{B,i}) }{ \sum(w_{A,i}) \sum(w_{B,i}) }.
    \label{eq:meth-cosine}
\end{equation}

The cosine similarity of $1$ indicates perfect match between events
and relative event counts, but does not require the same number of
total events.
Thus, the cosine similarity takes into account event counts in a more
flexible way than the weighted Jaccard score.  The cosine similarity can
be $1$ if the sets are of unequal sizes, as long as the relative
distribution of event weights is the same.

\subsubsection{Unweighted cosine similarity}
The unweighted cosine similarity is defined as
\begin{equation}
  C(A,B) = \frac{|A \cap B|}{\sqrt{ |A| |B| }}
\end{equation}
This is the analog of Eq.~(\ref{eq:meth-cosine}) when applied to
unweighted sets.  It has many of the same advantages and disadvantages
as the unweighted Jaccard score.

\subsection{The slicing process}
\label{sec:meth-time}
The slicing of the data is the core of the algorithm. %
It provides an efficient, one pass, linear time method of segmenting the groups of events in intervals %
where each interval $i$ comprises the time range within the
half-open interval $[t_{i}, t_{i+1})$.  The interval size is $\Delta t^*_i =
t_{i+1}-t_{i}$.  Our general procedure is:
\begin{enumerate}
\item Begin with some initial time $t_0$.  This is either the time of
  the very first event, or some specified time if one wishes to
  segment only a portion of the time period.
\item Find the optimal $\Delta t^*_0$ for the first interval.  Various
  values of $\Delta t$ are tried (see Sec.~\ref{sec:meth-dt}), and the
  optimum is the value which maximizes the similarity $J(\Delta t)$
  (see Sec.~\ref{sec:meth-peak}).  The interval is then set to $[t_0,
  t_0+\Delta t^*_0)$.
\item Repeat the previous steps until all data is treated, or until we reach a specified stop time.
  We go through time by setting the
  start of the next interval at the end of the previous, $t_i =
  t_{i-1}+\Delta t^*_{i-1}$.
\end{enumerate}

\subsubsection{Initial step: calculating the size of the first interval}
\label{sec:meth-firststep}
We begin with an initial time $t_0$, which is the lower bound of the
first interval.  If this is not provided by the user, it is the
time of the first event.  A test sequence of $\Delta t$s is generated
via one of the methods described in Sec~\ref{sec:meth-dt}.  For each $\Delta t$ generated, we compute
the intervals $ A(\Delta t) = [t_0, t_0+\Delta t)$ and $A'(\Delta t) =
[t_0+\Delta t, t_0+2\Delta t)$.  These will be the
next two intervals of width $\Delta t$ after $t_0$.  We then compute
our similarity score $J$ (Sec.~\ref{sec:meth-similarity}) between $A$ and
$A'$ as a function of $\Delta t$:
\begin{eqnarray}
  J(\Delta t) &=& J\left(A(\Delta t), A'(\Delta t)\right) \nonumber \\
              &=& J\left([t_0, t_0+\Delta t),  [t_0+\Delta t, t_0+2\Delta t)\right).
              \label{eq:J-max}
\end{eqnarray}
Eq.~(\ref{eq:J-max}) is
maximized as a function of $\Delta t$ to produce
$\Delta t^*$, our optimal interval size (see Sec.~\ref{sec:meth-peak} for a detailed explanation of this procedure). 
Once $\Delta t^*$ is found, the first interval $A$ is set to $[t_0,
t_0+\Delta t^*)$.  This interval is now fixed, the
starting time is updated to $t_1 = t_0+\Delta t_0$ and we proceed to the
propagation step.

\paragraph{Merging of the first two intervals}
\label{sec:meth-mergefirst}
This method provides an option which consists in merging the initial intervals. 
In this process, in the initial step, the first two intervals detected
($A$ and $A'$) are merged into one double-sized interval.  Continuing
from Sec.~\ref{sec:meth-firststep}, after $\Delta t^*$ is calculated,
the first interval $A$ is set to $[t_0, t_0+2\Delta t^*)$, and the new
starting time is set to $t_0 = t_0 + 2\Delta t$.  If this option is chosen, this process is done
at the beginning of the data slicing as well as after any critical events (which cause
a restart in the slicing process as explained in Sec.~\ref{sec:meth-critical}).
This avoids discarding the second interval $[t_0+\Delta t^*,
t_0+2\Delta t^*$, but causes the combined first interval to have a
different size distribution from subsequent calculations.

In cases where the data has only sharp transitions, this merge process
is advantageous, since the first two intervals will probably be more
similar than what comes after it.  However, in smoothly varying data,
merging should not be done because it changes the meaning of the
first interval relative to others.  This is noticeable when the first
interval is twice as long as others.  In the end, this decision must
be made with external information, and it should be used only if
needed.  Note that this decision is only relevant at the beginning of time,
when the method is first learning relevant timescales.

This is done in main text Fig.~1(a-c), where the data has sharp, clear
transitions.  Without this process, existing boundaries stay the same,
but each interval is divided into two.  Thus, there are pairs of $J=1$
intervals (self similar), followed by pairs of $J=0$ intervals
(critical events, Sec.~\ref{sec:meth-critical}).

\subsubsection{Propagation step}
\label{sec:meth-propagation}
Given our previous interval $A$ and starting time $t_i$ at the end of
$A$, we proceed to construct the next interval in a similar fashion 
to the initial step.  We generate our series
of $\Delta t$s and construct a series of intervals $B(\Delta t) = [t_i,
t_i+\Delta t)$ for each $\Delta t$.  Analogously to the initial step, we
compute the similarity score as a function of $\Delta t$,
\begin{eqnarray}
  J(\Delta t) &=& J\left(A, B(\Delta t)\right) \nonumber \\
              &=& J\left(A,  [t_i, t_i+\Delta t)\right).
              \label{eq:J-max2}
\end{eqnarray}
The difference to the initial step is that the first interval is
fixed, and only the second is changing.
We choose the $\Delta t^*$ which maximizes $J(\Delta t)$.  The next
interval is then fixed as $B = [t_i, t_i+\Delta t^*)$.

We repeat the propagation step indefinitely, until the intervals reach
the end of the data and all events are included.  For each iteration,
we take the $A$ as the previous interval, and begin at the next start
time $t_i = t_{i-1} + \Delta t^*_{i-1}$.

\subsection{$\Delta t$ generation}
\label{sec:meth-dt}
There are various methods to choose the $\Delta t$s to test in the
optimization process of the previous section.  We must explicitly
generate some values, because this is a numerical optimization.  It is important to do
this cleverly, or else the method can become very inefficient.  We
would rather not test every possible $\Delta t$ value, or test values
too far in the future.  We would prefer to check small $\Delta t$s
that are close together, but they should be spaced further apart at
long $\Delta t$.  We would rather check small $\Delta t$ values first,
since smaller intervals have fewer events to test, and thus are faster
to compute.

Also, there is a major opportunity for optimization.  As long as
$\Delta t$ increases, interval sets can be generated by simply adding
(via set union) new events (between $\Delta t_{n-1}$ and $\Delta
t_{n}$) to the previous interval set.  This is a very efficient
operation, and makes each individual iteration O($\Delta
t_\mathrm{max}$), assuming that $\Delta t$ only increases.

Since we do not have an \textit{a-priori} knowledge of the minimum or
maximum reasonable interval size, these are structured as generators
of $\Delta t$ values, returning an infinite sequence.  At a certain point,
the algorithm detects that we have searched enough, and that it is likely
that we already obtained the peak in similarity we were looking for,
which causes the generation of $\Delta t$ to stop
(as described in Sec.~\ref{sec:meth-stop-criteria}).

The methods described below are currently defined in the code.  Better
methods could be implemented in the future, including an actual
bisection to find the optimum.  Only the logarithmic method is fully
developed for actual use.

\subsubsection{Linear scan mode}
In this mode, the values $m+1d, m+2d, m+3d, ...$ are iterated.  The
parameter $d$ is the step size (1 by default), and $m$ is the minimum
step size (default to the same as $d$).  This method does not automatically adapt to
the data scale, thus reasonable values of step size and minimum step size must be provided.
Further,  this method is inefficient for data with a very long timescale, or data that evolves
in very different timescales.

\subsubsection{Logarithmic scan mode}
In this mode, we begin with a base scale of
\begin{equation}
  m = 10^{\left\lfloor \log_{10}(t_e-t_i)  \right\rfloor}
\end{equation}
where $t_e-t_i$ is the time to the next event after the interval start
time $t_i$.
Thus, $m$ is the greatest power of 10 less than the time to the next
event.  This allows the scan mode to adapt to the actual data sizes.
Then, we return the sequence of values:
\begin{equation}
  1m, 2m, \ldots, 99m, 100m, 110m, 120m, \ldots, 990m, 1000m,
  1100m, \ldots, 
\end{equation}
which produces a logarithmic time scale with reasonable precision at
all points.

\subsubsection{Event-based scan mode}
In this mode, for every distinct event time $t_e > t_i$, we return the
corresponding $\Delta t = t_e-t_i$.  Recall that $t_i$ is the start of
the interval.  In this way, every time step with presence of events
 is tested.  This mode offers the greater precision in aligning
intervals with events, but can be inefficient for very large datasets
at very long times, where each individual event is unlikely to have
an effect on the Jaccard score.

\subsubsection{An ideal mode}
An ideal method would combine parts of the above methods.  It would
begin with a logarithmic scanning, but ideally with some fixed
multiplier such as $1.01$.  The next time is found by $\Delta
t_\mathrm{next} = \lceil 1.1\Delta t \rceil$.  Here, the ceiling
operator $\lceil \cdot \rceil$ means \textit{the time of the next
  event equal to or after the given time}.  This allows a logarithmic
increase in time, while always aligning with actual events and
skipping non-present events.  After a maximum is found, we would
backfill with a bisection algorithm to find the exact event which
produces a global optimum for the similarity peak.

The downside to this method is that it requires many searches through
our data to find the event-ceiling.  This is implemented as a fast
database search, but still requires extra operations.  Also, the
bisection stage would ideally need to be able to increment sets both
forward and backwards in time.  Currently, the process of set construction is
optimized to incrementally build up the sets while moving forward in
$\Delta t$.  Going backwards is possible with weighted sets (though 
this procedure needs to be done with caution, watching out for floating point errors), 
but with unweighted
sets this operation is not possible.  This biases us to do a more thorough scan
going only forward in time.  The current logarithmic implementation is seen as a
good trade-off between simplicity, accuracy, generality, and
computational performance.

\subsubsection{Criteria for stopping the $\Delta t$ generation}
\label{sec:meth-stop-criteria}
The above methods do not specify when to stop searching new $\Delta t$ values 
(except when using greedy maximum finding, as we will see in Sec.~\ref{sec:greedy}).  
The intuition is that there will always be some maximum, independently of the
number of values of $\Delta t$ that we choose to scan. In order to have a good chance
of finding the global optimum, we need to carefully adjust how many time steps we will search 
after the latest found peak. If we decide to search in a small time window, we have great
chances to get trapped in a local maxima. Instead, if we opt for scanning a very large time window, 
we will certainly sacrifice the computational efficiency. As usual, a balance between these two 
opposite cases is desired. We achieve this by continuing to scan until we have tested all $\Delta t <
25\Delta t^*$ and $\Delta t$ less than 25 times the previous round's
$\Delta t^*$, if there is a previous interval. As mentioned, the multiplier can be adjusted lower for
better performance or higher for less risk of missing future peaks,
but this is the subject of further research.

\subsection{Maximum finding}
\label{sec:meth-peak}
The $\Delta t$ values are given by one of the methods from
Sec.~\ref{sec:meth-dt}, and we wish to find the optimum value of $\Delta
t^*$ such that it maximizes  $J(\Delta t)$.  %
The first consideration is that there will likely be many local maxima.
Some will be caused by general fluctuations in the data.  However, when scanning on
a larger scale, we may see different local maxima, which can be
interesting in their own right, because they may indicate different time
scales of the system.  These can be seen by plotting $J(\Delta t)$
as a function of $\Delta t$, primarily for the first interval.
Sec.~\ref{sec:peak-factor} discusses some methods of adjusting the
timescale detection to find other maxima.

Then the question is under what conditions do we expect non-trivial maxima to exist.  
When there is a long-term evolution of the system (events active at $-\infty$ are
different from those at $+\infty$), then we will obtain a decrease in similarity as 
$\Delta t$ becomes longer. In this case, there will not be a maximum.
However, if there is not a long-term evolution (i.e.\ for each event ID,
that event has the same probability of occurring at any point in
time), then the similarity $J(\Delta t)$ may continually increase.
In this case, the basic assumptions of our method are violated, but the answer is
correct anyway: as all times are statistically self-similar,
we expect one giant interval covering all time because there are no
sub-divisions of distinct character.

In brief, the process of maximum finding works as follows. 
For each $\Delta t$, we calculate the
similarity $J(\Delta t)$.  We search for the maximum value of
$J(\Delta t)$ in an on-line fashion.  The $\Delta t$ iteration
(see Sec.~\ref{sec:meth-dt}) produces a continuous stream of $\Delta t$
values.  After each $\Delta t$ is produced, $J(\Delta t)$ is
calculated (as in Sec.~\ref{sec:meth-similarity}).  Then, the list of all
$\Delta t$ and $J(\Delta t)$ are examined by the methods that we will present next in this 
section, which return $\Delta t^*$, the optimum interval
size.  This $\Delta t^*$ is fed back to the stop criteria in
Sec.~\ref{sec:meth-stop-criteria} and used to decide when we should
cease exploring further $\Delta t$.  Once the stop criteria is met,
the iteration stops and the final $\Delta t^*$ is known.

\subsubsection{Longest}

This method is our standard approach.  We search for the maximum
similarity value of $J(\Delta t)$ from all possible $\Delta t$ values.
If there are multiple values of $\Delta t$ with the same maximum, we
pick the greatest one as $\Delta t^*$.

\subsubsection{Shortest}

This is similar to ``Longest'', but choosing the shortest $\Delta t$.

\subsubsection{Greedy}
\label{sec:greedy}
In this method, as soon as $J(\Delta t)$ decreases for the first time, we stop testing
other values.  This method is much more efficient in terms of computation
time.  While there is usually one clear peak, there are often local
fluctuations which cause this method to give a local maximum far
before the global maximum.  If the number of events is large enough to
reduce the effects of fluctuations, or if the highest computational
performance is needed for streaming data, this method could be useful.
If this method is used, the stop criteria of
Sec.~\ref{sec:meth-stop-criteria} is unneeded.

\subsubsection{Detecting other timescales}
\label{sec:peak-factor}
One of the basic assumptions of this method is that the similarity score
increases, and then decreases.  We would hope that as the similarity
starts going down, we can notice a peak.  The peak finders, as
described above, do not find multiple local maxima.  This is because
choosing a peak requires some heuristic for how significant of a peak
should be detected.  The two extremes are represented by the
``longest''/``shortest'' peak finders (scan forward in time as much as
possible, find peak as global maximum in scanned area) and ``greedy''
(stop at the first decrease, as soon as any maximum is found; in other
words the first local maximum).  This corresponds to finding the
longest similar time scales and shortest similar time scales.

\textbf{Peak factor.}
The ``peak factor'' is a way to balance these.  It is implemented only
for the ``longest'' and ``shortest'' maximum finding methods.  Using
this option, once $J(\Delta t_\mathrm{max})$ drops to less than
$\mathrm{peak\_factor}\times J(\Delta t^*)$, then we stop searching
any longer times and return $\Delta t^*$.  This provides a way to
break out of the search after the first peak.

\textbf{Pastpeak factor.}
The factor of 25 present in Sec.~\ref{sec:meth-stop-criteria} can be
adjusted.  This limits the forward search time and provides another
way to limit the timescale searched.

\textbf{Pastpeak max, pastpeak min, search min.}
One can directly set the time scales searched.  One can set a maximum
or minimum amount of time to search past $\Delta t^*$ (as in
Sec.~\ref{sec:meth-stop-criteria}).  One can also set a minimum amount
of time to search past $\Delta t=0$.  This may be useful to avoid the
effects of extreme fluctuations in $J$ at small $\Delta t$.  Unlike
the other options, these require a pre-existing knowledge of likely
time scales within the data.  Because of this, these options are not
used in this work or enabled by default.  The existing options work to
achieve the goal of detecting the longest possible timescales without
needing any parameters.

\subsubsection{Critical event detection}
\label{sec:meth-critical}
At some times (critical times), the character of the active events in the entire data
changes instantly.  When this happens, comparison with the previous interval will give bad
results, since similarity is by definition going to always be low.
Similarity scores could also remain zero, if there is no overlap in the set of 
events before and after the critical time.  Thus, the signature of critical events is
low but continually increasing similarities with no peak found. At this point, the
algorithm will tend to produce longer and longer intervals,
pointlessly trying to maximize similarity and eventually return
too large intervals.   At this point, using the ``initial step'' process
(see Sec.~\ref{sec:meth-firststep}) is better than using the propagation
step (see Sec.~\ref{sec:meth-propagation}).

Using the ``initial step'' process means that once a critical event occurs, 
the previous interval is forgotten and segmentation restarts as in Sec.~\ref{sec:meth-firststep}.  
To decide whether an event can be considered critical, we require that 
i) the similarity corresponding to the last $\Delta t$ is greater than 0.95 of the peak similarity
and ii) we have not reached the end of the dataset.

An example of critical events can be seen in main text Fig.~1(a-c).
Each interval boundary occurs at a critical event.

\section{Dataset descriptions}
\label{sec:datasets}

\subsection{Periodic toy model}
This model creates data with evolving timescales while having a uniform event rate. 
The latter means that at every point in time, there is the same expected value of the number of events.
Our method does not group the data in intervals using the overall event rate, 
it detects intervals based on the \textit{identity} of events at a given time.  
In this model, the characteristic active events change slowly over time, but at a varying
rate.  This adds the ``evolutionary timescales'' that our method detects.
When the timescales evolve faster, intervals are shorter, and
when timescales evolve slower, intervals are longer. 

In this periodic toy model,
a universe of $N=1000$ event IDs are created.  Upon initialization,
each event is independently placed into an activated status with
probability $q$.  At each timestep, each active ID emits an event
with probability $p$.  This $p$ produces the short-term repetition, on
a timescale of $1/p$.
Also, at each timestep (and before event
creation), a fraction $c(t)$ of event IDs are picked.  Each of these
IDs has its active status updated, being made active with probability
$q$ regardless of its previous state.  This provides the long-term
change: on average, each active event tends to lasting on the order of
$1/c(t)$ time steps until it is deactivated.
This preserves the mean total
number of activated IDs as $qN$ at all times.  At each timestep, on
average $pqN$ events are emitted.

The changeover rate $c(t)$ is periodic, however, obeying
\begin{equation}
  \label{eq:periodicct}
  c(t) = c_0 + c \left(\frac{1}{2} - \frac{1}{2} \cos\left({2\pi t \over \tau}\right)\right),
\end{equation}
with $c_0$ being the minimum changeover rate and, $c$ being the scale
of changeover, and $\tau$ being the period.
Thus, at some points, the IDs of the expressed
events is changing more rapidly than at other times, and this is the
cause of longer and shorter intervals.

When there is a critical event at time $t_\mathrm{crit}$, we
exceptionally set $c(t_\mathrm{crit})=1$ at that time.  This produces
a state of the system decorrelated from the previous time.  According
to the model, this changeover occurs before the timestep, which
matches with placing a new interval at $t_\mathrm{crit}$ in our
half-open segment convention.  However, the total universe of event
IDs stays the same, and there is no statically observable change in
behavior.

In this work, we use $N=1000$, $\tau=500$, $p=0.2$, $q=0.2$, $c_0=0$, and
$c=0.01$.

\subsection{Enron}
The Enron Email Dataset consists of emails of approximately 150 senior
managers of the Enron Corporation, which collapsed in 2001 after
market manipulation was uncovered, leading to an accounting scandal
\cite{klimt2004enron}.  This dataset was made public during the legal
investigation concerning the Enron Corporation, which ended up with
the bankruptcy and collapse of this corporation. Here, each event is
an email communication sent or received by any of the senior managers
who were subsequently investigated. During the time recorded in the
dataset, there were major structural changes in the company,
executives were hired and fired and new products were launched. This
exogenous information was sometimes also reflected in the data as
changes in volume and in the identities of active events.

\begin{figure*}
   \centering
   \includegraphics[width=2\columnwidth]{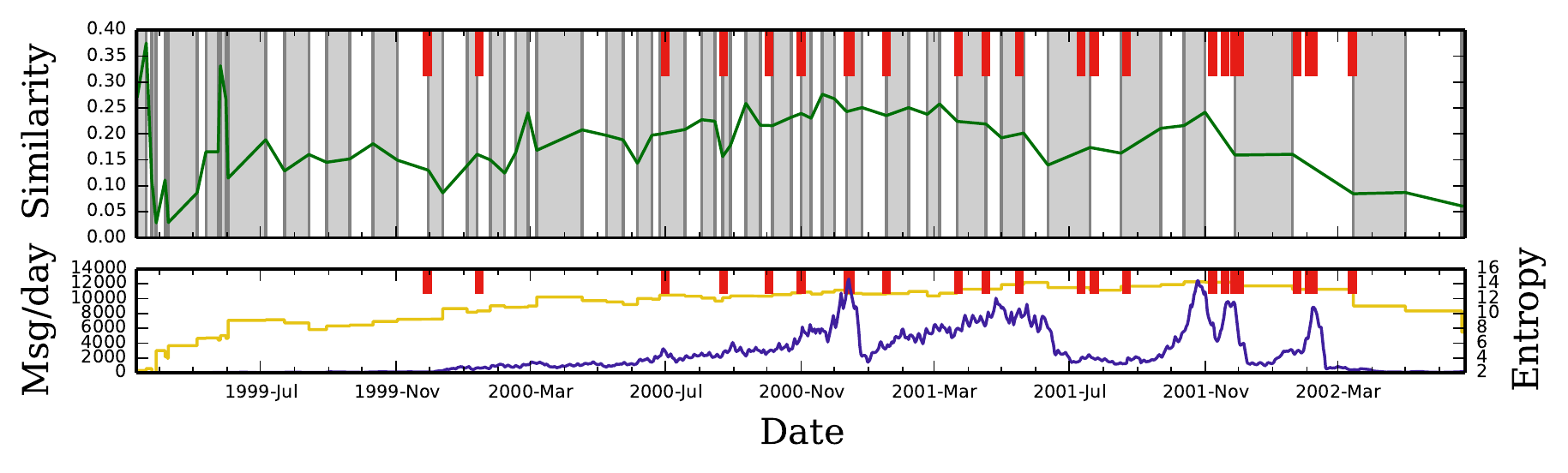}
   \caption{Slicing of the full Enron dataset.  The format of this figure
     is as in Fig. 2 %
     from the main text. Thick red
     bars at the top of the plots correspond to actual dates of major
     events.  This is the slicing of the whole dataset,
     not just the links between executives. %
     \label{fig:enron-full}}
 \end{figure*}

We have all bidirectional email communication
to and from each key person at a resolution of one day.  An event is
the unordered pair (source, destination) of each email.  Our
particular input data is already aggregated by day, therefore each
event is only present once each day, with a weight of the number of
mails that day.
Table~\ref{tab:enron-stats}-\ref{tab:enron-stats-full} list the basic properties and
Table~\ref{tab:enron} lists the major events of this dataset. In Fig.~\ref{fig:enron-full}, we show the full Enron data.  This
includes one event for every message sent, even to other executives.

\begin{table}
  \centering
  \begin{tabular}{l|c} \hline
    Number of events             & 43599 \\
    Number of distinct events    & 1275 \\
    First event time             & 1999-05-11 \\
    Last event time              & 2002-06-21 \\
    Total count/weight of events & 43599 \\
  \end{tabular}
  \caption{Basic properties of Enron dataset (core network)}
  \label{tab:enron-stats}
\end{table}

\begin{table}
  \centering
  \begin{tabular}{l|c} \hline
    Number of events             & 802481 \\
    Number of distinct events    & 207071 \\
    First event time             & 1998-05-26 \\
    Last event time              & 2002-07-11 \\
    Total count/weight of events & 2933183 \\
  \end{tabular}
  \caption{Basic properties of Enron dataset (full data)}
  \label{tab:enron-stats-full}
\end{table}

\begin{table}
  \centering
  \begin{tabular}{l|l}
    Date       & Event \\ \hline
    1999-11-29 & Launch of enron online    \\
    2000-01-15 & Launch of EBS    \\
    2000-07-01 & EBS-blockbusters partnership    \\
    2000-08-23 & Stock all-time high    \\
    2000-10-03 & Enron attorney discusses Belden's strategies    \\
    2000-11-01 & FERC exonerates Enron    \\
    2000-12-15 & EBS \$53m `profit'     \\
    2000-12-13 & Skilling announced to be CEO    \\
    2001-01-17 & Blackouts in CA    \\
    2001-03-23 & Enron schedules conference call to boost stock    \\
    2001-04-17 & The `asshole' call    \\
    2001-05-17 & Schwarzenegger, Lay, Milken meeting    \\
    2001-07-12 & Quarterly conference call    \\
    2001-07-24 & Skilling meets with analysists and investors in NY    \\
    2001-08-22 & Watkins raises accounting irregularities    \\
    2001-11-08 & Dynegy agrees to buy enron    \\
    2001-11-19 & Enron restates its third quarter earning    \\
    2001-11-28 & Enron shares plunge below \$1    \\
    2001-11-28 & Dynergy deal collapses    \\
    2001-12-02 & Enron files for bankrupcy    \\
    2002-01-23 & Stephen Cooper takes over as Enron CEO    \\
    2002-02-03 & Lay cancels Senate committee appearance    \\
    2002-02-07 & Fastow, Kopper, Lay invoke the Fifth    \\
    2002-02-07 & Skilling and Watkins testify    \\
    2002-03-14 & Arthur Anderen LLP indicted
  \end{tabular}
  \caption{Major events of the Enron collapse}
  \label{tab:enron}
\end{table}

\subsubsection{Comparison with the \textit{cpdetect} method}  
The \textit{cpdetect} method is run as described in Ref.~\cite{peel14}.  A part of this process is to
pre-aggregate all data into windows of one week size, and thus any
changepoints are \textit{a priori} limited in their accuracy.  This
process is not necessary in our work, and one possible use of our method is
providing initial intervals for other methods such as cpdetect.

The output from our method and cpdetect is not directly comparable.
The cpdetect method produces only significant changepoints, while our
method should produce snapshots distributed throughout the entire time
period.  However, we would expect that the similarity $J$ is lower
than average at each changepoint.  If so, it will indicate that our
method can show at least the major underlying dynamics of the system.
We show this with a hypothesis test.  To do this, calculate a test
statistic which is our mean (interpolated) similarity value
corresponding to each change point found by cpdetect.  We randomize
the times of change points by picking times uniformly from our entire
time range of changepoints, which shuffles the change point sequence.
We do this $1\times10^4$ times to compute a distribution of our test
statistic in the random case.  We find that our actual value of $X$ is
significant at $p=0.0033$. This means that our choice of similarity
does correlate with the evolutionary timescale of the system. If one
looks at low values of $J$, one can find changepoints, and our process
of maximizing $J$ finds self-similar intervals.

\subsection{Reality mining dataset}
The Reality Mining dataset covers a group of persons affiliated with
the MIT Media Lab who were given phones which tracked other devices in
close proximity via the Bluetooth personal area network protocol
\cite{eagle2006reality}.  We say an event is every ordered pair of 
(personal\_device, other\_device).  We only have data from the
personal devices of the 91 subjects who completed the experiment, not
full ego networks.  The data contains 1881152 unique readings from
2004 August 8 until 2005 July 14, being most of them centered in the middle of
the academic year.  Table~\ref{tab:reality-stats} lists basic
properties and Table~\ref{tab:reality} lists the major events of this
dataset.  The major events are from the MIT academic calendar or
internal Media Lab events, with which most subjects were affiliated
\cite{MITcalendar}.

As expected, we find many more segment intervals than major events. In
the slicing, we can see that in the first weeks, where students are
still relocating and meeting new people, the change in the events'
identities is fast, thus the intervals are shorter.  On the other hand,
the interactions throughout the rest of the academic year are pretty
stable, and the intervals are sized accordingly.  Additionally, we see
that several key events are detected.  The major Thanksgiving holiday
(Nov 25) is detected a day early, as people begin traveling. There are
intervals before and after the start of the December exam week, and
then a stable period after that when people are on holiday until the
university opens again (Jan 3).
After that, we see segment boundaries that exactly align with the
beginning of classes (Feb 1), the beginning of spring break (Mar 19),
and the last day of classes (May 12). All of this can be detected
despite the fact that many subjects are research staff, who are not
bound by any academic calendar.

\begin{table}
  \centering
  \begin{tabular}{l|c} \hline
    Number of events             & 1881152 \\
    Number of distinct events    & 51466 \\
    First event time             & 2004-08-01 00:26:38 \\
    Last event time              & 2005-07-14 17:41:19 \\
    Total count/weight of events & 1881152 \\
  \end{tabular}
  \caption{Basic properties of the Reality Mining dataset}
  \label{tab:reality-stats}
\end{table}

\begin{table}
\centering
\begin{tabular}{l|l}
Date       & Event \\ \hline
2004-09-08 & First day of classes    \\
2004-10-18 & Sponsor week end    \\
2004-10-25 & Sponsor week start    \\
2004-11-08 & Exam week start    \\
2004-11-12 & Exam week end    \\
2004-11-25 & First day of thanksgiving    \\
2004-12-09 & Last day of classes    \\
2004-12-13 & Exams start (4 days)    \\
2005-01-03 & Independent activities period start    \\
2005-02-01 & Start of spring classes    \\
2005-03-19 & Start of spring break    \\
2005-03-28 & End spring break    \\
2005-05-12 & Last day of classes    \\
2005-05-16 & Exams start    \\
2005-06-03 & Commencement
\end{tabular}
  \caption{Major events of the Enron dataset}
  \label{tab:reality}
\end{table}

\subsection{UCL hashtags}
In this dataset, we scrape Twitter for all hashtags containing
\texttt{\#UCL}, referring to the UEFA (European) Champions League
tournament.  We scrape between the dates of 30 April 2015 and 08 May
2015, covering the first leg of the two semifinal games of the tournament.  An event is
any hashtag co-occurring with \texttt{\#UCL}, except \texttt{\#UCL}
itself.  If there are multiple hashtags in the same tweet, each counts
as one event.  Thus, events (hashtags) reflect what people are talking
about in conjunction with the tournament.  All hashtags are
interpreted as UTF-8 Unicode and case-normalized (converted to
lowercase) before turning into events.  Table~\ref{tab:hashtags-stats}
lists basic properties and Table~\ref{tab:hashtags} lists the major
events of this dataset.

In the slicing (main text Fig.~5(a)), we see much intervals
segments during the games of May~5 and May~6 than in the rest of the dataset. This corresponds to the
must faster turnover of interesting topics during the game. In
addition, we can spot the games that all four teams played in their
respective national championships on the previous weekend of May~2-3,
which trigger as well some discussion on the forthcoming Champions
League matches. 
If we focus our attention on the first of the two games (main text Fig.~5(b)), we observe 
a shorter segment size at the beginning of the game and during the three
goals at times 20:53 (Juventus), 21:12 (Real Madrid), and 21:58
(Juventus), compared to the rest of the match. 
In particular, we notice a more noticeable effect when the second
half started with the tied game at 21:45, and when the third goal is
scored, because it was a penalty kick which took Juventus to the
lead.%
Game 2 was a lopsided 3-0 victory for Barcelona against Bayern Munich.
Given the one-sided game, the analysis shows a fairly uniform pattern
throughout the game, until the goals are scored at 22:17, 22:20, and
22:34. The game ends right after the last goal.  In the initial part
of the game, the dynamics continually slowed down as time goes on
without major events.  Immediately after the game, the audience reacts
much more strongly than in the last game, probably because at that
point one starts speculating about the possible finalists of the
competition.

\begin{table*}
  \centering
  \begin{tabular}{l|c|c|c} \hline
                                 & Entire data & Game 1 & Game 2 \\ \hline
    Number of events             & 401849             & 87900              & 120123 \\
    Number of distinct events    & 8650               & 2660               & 3161 \\
    First event time             & 2015-04-30 16:08:42& 2015-05-05 19:45:00& 2015-05-06 19:45:00 \\
    Last event time              & 2015-05-08 01:59:45& 2015-05-05 23:44:59& 2015-05-06 23:44:59 \\
    Total count/weight of events & 401849             & 87900              & 120123 \\
  \end{tabular}
  \caption{Basic properties of the UCL dataset.  All times are in
    CEST, the local timezone of the matches.}
  \label{tab:hashtags-stats}
\end{table*}

\begin{table}
  \centering
  \begin{tabular}{l|l}
    Date             & Event \\ \hline
    2015-05-05 20:45 & Game 1 begins \\
    2015-05-05 20:50 & Yellow card for Bonucci (Juventus) \\
    2015-05-05 20:53 & Goal by Morata (Juventus) \\
    2015-05-05 21:12 & Goal by Ronaldo (Real Madrid) \\
    2015-05-05 21:29 & First half ends \\
    2015-05-05 21:45 & Second half begins \\
    2015-05-05 21:58 & Goal by T\'evez (Juventus) \\
    2015-05-05 22:33 & Game 1 ends \\
    2015-05-06 20:45 & Game 2 begins \\
    2015-05-06 21:29 & First half ends \\
    2015-05-06 21:45 & Second half begins \\
    2015-05-06 22:17 & Messi (Barcelona) scores \\
    2015-05-06 22:20 & Messi (Barcelona) scores \\
    2015-05-06 22:34 & Neymar (Barcelona) scores \\
    2015-05-06 22:34 & Game 2 ends
  \end{tabular}
  \caption{Major events of the first leg of the Champion's League semifinals}
  \label{tab:hashtags}
\end{table}

\begin{figure*}
   \includegraphics[width=2\columnwidth]{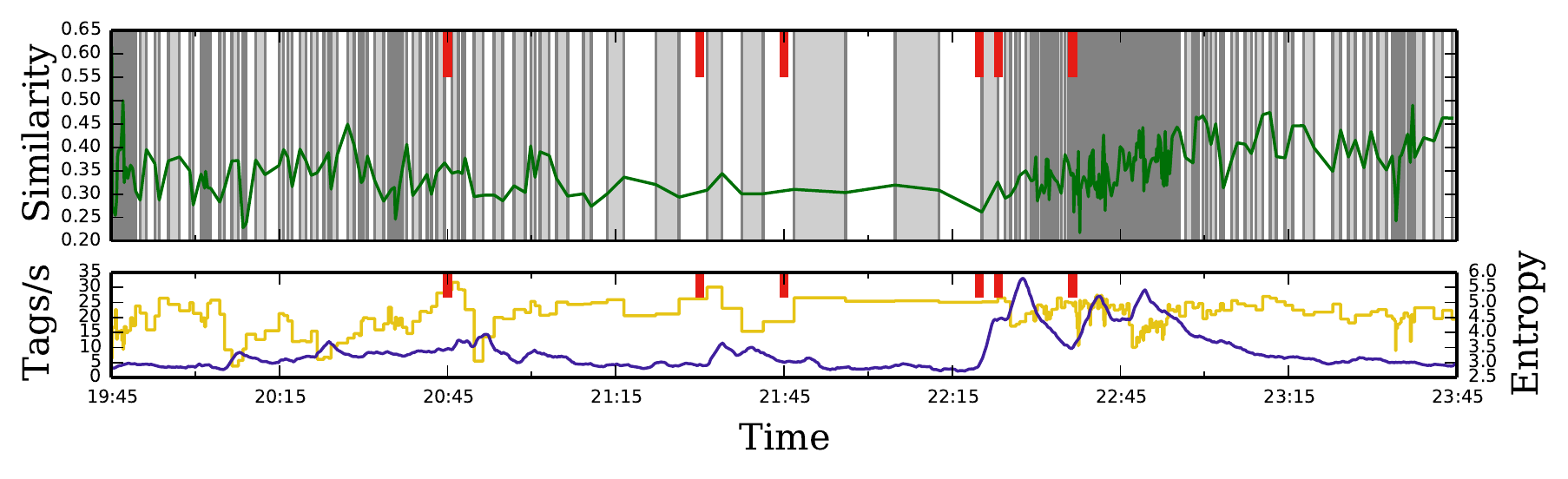}
   \caption{Second semifinal of the UEFA Champions League 2014-2015,
     between Barcelona and Bayern Munich.  Game times are the same as
     in main text Fig.~5(b) in the main text.  This game was
     one-sided with goals made by only one team.  Consequently,
     internal structure for the slicing is relatively uniform and
     slowing down with time, especially since all goals were scored
     within the last 20 minutes of the game.  Nevertheless, we can
     detect interval boundaries for halftime, and after the three
     goals are scored (22:17, 22:20, and 22:34), we have a very high
     turnover rate of discussion topics.}
   \label{fig:hashtags2}
 \end{figure*}

The detail of the second game is presented in Fig.~\ref{fig:hashtags2}.
There is a flurry of activity at the start, but as the game progresses
without goals dynamics slow down.  At the end of the game, there are
three goals (beginning at 22:17).  At this point, the system becomes
extremely active with a very diverse and rapidly changing topics, and
intervals are very short for a while.

\end{document}